\documentclass[twocolumn,prl,aps,showpacs]{revtex4}
\usepackage{epsfig,latexsym,graphicx}

\def \ba {\begin{eqnarray}}
\def \ea {\end{eqnarray}}

\def \ua {\uparrow}
\def \da {\downarrow}

\begin{document}

\title{Tunneling spectroscopy of
$s\pi$ pairing state \\ as a model for FeAs superconductors }

\author{Han-Yong Choi} \affiliation{Department of Physics and
Institute for Basic Science Research, SungKyunKwan University,
Suwon 440-746, Korea.}

\author{Yunkyu Bang}\affiliation{Department of Physics, Chonnam National
University, Kwangju 500-757, Korea. \\
Asia Pacific Center for Theoretical Physics, Pohang 790-784,
Korea.}

\begin{abstract}
We present the self-consistent Bogoliubov-de Gennes calculations
of an $s\pi$ pairing state of two band superconductivity as a
model for the FeAs superconductors. The $s\pi$ state is an
$s$-wave pairing state with an internal $\pi$ phase, that is,
nodeless gaps on each band but with the opposite sign. The novel
features of this state are investigated by calculating the local
density of states of the $\pi$ phase superconductor/normal metal
bilayers. Because of the sign reversal between the two
condensates, the zero bias conductance peak appears as observed
in tunneling spectroscopy experiments on FeAs superconductors.
This eliminates the major obstacle to establish the $s\pi$ state
as the pairing symmetry of the FeAs superconductors.

\end{abstract}



\maketitle

$Introduction$ -- The recent discovery of the iron based pnictides
superconductors generated enormous interests in the community
\cite{YKamihara06,YKamihara08,PhysToday08,GFChen08,XHChen08}.
Like the cuprate superconductors, the pnictides are highly
two-dimensional, the undoped parent materials are in the
antiferromagnetic state, and superconductivity emerges when the
antiferromagnetism is suppressed \cite{Cruz08} upon doping either
electrons or holes \cite{Rotter08} into the FeAs planes. The
urgent question is the pairing mechanism of the FeAs
superconductors. This will hopefully pave the way to coherently
understand the still elusive pairing mechanism of the high
temperature superconductivity. The first step towards this goal
is to establish the pairing symmetry of the FeAs superconductors.

Conflicting experimental results are being reported with regard
to the orbital pairing symmetry. Some experiments are
unambiguously seeing a fully gapped nodeless superconducting
state, while others point towards a gap with nodes: The several
independent ARPES experiments on single crystals reported the
full gaps around the hole Fermi surface although the gap feature
is less clear around the electron Fermi surface
\cite{LZhao08,HDing08,TKondo08}. The temperature dependence of
the penetration depth strongly suggests a nodeless gap structure
\cite{CMartin08}. The infrared spectroscopy observed the
superconductivity induced features which were best described in
terms of $s$-wave pairing \cite{GLi08}.

The evidences for nodes are also accumulating: The nuclear
spin-lattice relaxation rate $1/T_1$ clearly showed no coherence
peak and $\sim T^3$ dependence below $T_c$
\cite{KMatano08,Grafe08,Nakai08,Mukuda08}. Also, the resonant
spin excitations were observed by neutron scattering experiments
with the momentum transfer corresponding to the magnetic order
and the energy transfer consistent with the $T_c$ scaling
\cite{Christianson08}. These observations strongly suggest a gap
with nodes. On the other hand, the evidence for the singlet
pairing and multiple gaps seems convincing from the Knight shift
and other thermodynamic measurements \cite{KMatano08,Grafe08}.

The various tunneling spectroscopy, however, seems confusing. The
Andreev spectroscopy reoprted by Chen and others \cite{TYChen08}
was well fitted by the $s$-wave Blonder-Tinkham-Klapwijk
formalism \cite{Blonder82}, although they observed zero bias peak
(ZBP) which seemed difficult to understand within the $s$-wave
pairing formulation. Other point contact and tunneling
spectroscopy measurements also reported the ZBP which strongly
suggests a sign change of the pairing order parameter
\cite{YWang08,LShan08,Millo08,Samuely08}.

Many ideas have been put forward to understand the seemingly
conflicting experimental observations on the FeAs materials. Among
them, particularly appealing is the sign changing $s$-wave
pairing state as advanced by Mazin and coworkers \cite{Mazin08}.
It seems to be able to explain the experimental observations
indicating the full gap as well as a gap with nodes
\cite{Parker08,Chubukov08,Kuroki08}. It was noticed early on that
there exists this type of solution to a multi-band BCS gap
equation \cite{Aronov73,Rice91}. It is quite exciting that it
seems to be actually realized in the pnictides. We adapt this
proposal in this Letter, and call it ``$s\pi$'' pairing state
because of its natural connection with other $\pi$ states. It is
interesting to note the similarity between the internal $\pi$
state of the $s\pi$ pairing and the ``external'' $\pi$ state of
the superconductor/ferromagnet (S/F) bilayers. It is well
established that a Cooper pair in the F side of an S/F bilayer
picks up a non-zero center of mass momentum and the singlet
pairing order parameter oscillates as a function of position in
the F side \cite{Demler97}. The superconductivity induced
suppression of local density of states (LDOS) in the subgap
energy range becomes enhanced where the order parameter changes
its sign \cite{Kontos01}. Similarly, the internal $\pi$ state is
expected to exhibit the zero bias conductance enhancement in the
subgap region \cite{index}.

We will show that it is indeed the case. This eliminates the major
obstacle to establish the $s\pi$ state as the pairing symmetry of
the FeAs superconductors. We solve the Bogoliubov-de Gennes (BdG)
equation self-consistently to calculate LDOS of the $\pi$ state
superconductor/normal metal (S/N) bilayers. It will be
demonstrated that the observed zero bias conductance peaks which
seemed difficult to understand with a fully gapped pairing state
may be consistently understood within the $s\pi$ pairing state.

$Model$ -- The FeAs superconductors have the disconnected electron
and hole Fermi surfaces. A minimal model has to include two
bands; a hole band around the $\Gamma=(0,0)$ point and an electron
Fermi surface around the $M=(\pi,\pi)$ point \cite{Raghu08}. We
write the BdG equation for S/N bilayers with two band
superconductivity as
 \ba
 \label{bdg2}
H &=& H_S +H_N +H_{int}, \\
 \label{super}
 H_S &=& \sum_{n,k,\sigma}\sum_{1\le y \le y_S} \xi_{n k} c_{nk\sigma}^\dag (y)
c_{nk\sigma} (y) \\ \nonumber
 &-& \sum_{n,k}\sum_{1\le y \le y_S} \left[ \Delta_{nk}(y) c_{nk\ua}^\dag (y)
c_{n,-k\da}^\dag (y) + h.c. \right] \\ \nonumber
 &-& t_0 \sum_{n,k,\sigma}\sum_{1\le y \le y_S} \left[ c_{nk\sigma}^\dag (y)
c_{nk\sigma} (y+1) + h.c. \right], \\
 H_N &=& \sum_{k,\sigma}\sum_{y_S+1\le y \le y_t} \xi_{k} a_{k\sigma}^\dag (y) a_{k\sigma}
(y) \\ \nonumber &-& t_0 \sum_{k,\sigma}\sum_{y_S+1\le y \le y_t}
\left[ a_{k\sigma}^\dag
(y) a_{k\sigma} (y+1) + h.c. \right] , \\
 H_{int} &=& -t_1 \sum_{k,\sigma} \left[ c_{1k\sigma}^\dag (y_S) a_{k\sigma} (y_S+1)
+ h.c. \right] \\ \nonumber
 &-& t_2 \sum_{k,\sigma} \left[ c_{2k\sigma}^\dag (y_S) a_{k\sigma} (y_S+1)
+ h.c. \right].
 \ea
Here, the subscript $n=1$ and 2 refer to the hole and electron
Fermi surfaces around the $\Gamma$ and $M$ points in the momentum
space, respectively. $k$ is the intra-layer crystal momentum in
the $z$-$x$ plane. $y$ is perpendicular to the interface between S
and N. $y_S$ and $y_N$ are thickness of S and N layers,
respectively, in the unit of the distance between neighboring
single layers, and $y_t=y_S+y_N$. $a$, $c_1$, and $c_2$ are the
electron operators for the N, the hole band, and the electron
band of S, respectively. The self-consistency relation for the
gap function $\Delta_{nk}(y)$ is given by
 \ba
 \label{delta}
\Delta_{nk}(y)= \sum_{n',k'}V(n,k;n',k') \left< c_{n',-k'\da}(y)
c_{n'k'\ua}(y) \right>,
 \ea
where $V(n,k;n',k')=V_{nn'}$ is the pairing interaction.

Let us first consider bulk $s$-wave two band BCS
superconductivity. Putting $V_{ij}N_j=\lambda_{ij}$, where $N_1$
and $N_2$ are the DOS per spin at the Fermi level for the band 1
and 2, respectively, Eqs.\ (\ref{super}) and (\ref{delta}) are
reduced to \cite{Suhl59,Rice91}
 \ba
 \label{twogap}
\Delta_1 = \lambda_{11} F(\Delta_1)\Delta_1 +\lambda_{12} F(\Delta_2)\Delta_2, \\
 \nonumber
\Delta_2 = \lambda_{21} F(\Delta_1)\Delta_1 +\lambda_{22}
F(\Delta_2)\Delta_2,
 \ea
where $F(\Delta)$ is given by
 \ba
F(\Delta) = \int_0^{\omega_D} d\xi \ \frac{1}{E}
\tanh\left(\frac12 \beta E \right),~~E=\sqrt{\xi^2 +\Delta^2}.
 \ea
At $T=T_c$, $F=\ln(1.14\omega_D /T_c)$. The $T_c$ is the highest
temperature where the larger eigenvalue of Eq.\ (\ref{twogap})
becomes 1.

To see physics through more clearly, consider the simplest case of
$\lambda_{11}=\lambda_{22}=\lambda$ and
$\lambda_{12}=\lambda_{21}=\lambda'$. The single band BCS
expression for the critical temperature $T_c = 1.14 \omega_D
e^{-1/\lambda}$ is now replaced by one of the two expressions:
 \ba \label{plus}
T_c = 1.14 \omega_D e^{-1/(\lambda+\lambda')},\\ \label{negative}
 T_c = 1.14 \omega_D e^{-1/(\lambda-\lambda')},
 \ea
For $\lambda'<0$, Eq.\ (\ref{negative}) is the appropriate
expression, and the pairing order parameter $\Delta_1$ and
$\Delta_2$ on the two bands acquire the $\pi$ phase shift, which
is the case considered here. The negative pairing interaction in
one band BCS theory does $not$ permit superconductivity. For the
two band case, however, the negative interaction is turned to
induce pairing by generating the sign reversal between $\Delta_1$
and $\Delta_2$ as can easily be seen from Eq.\ (\ref{twogap}). The
physical nature of the negative interaction parameterized in
terms of $\lambda'$ we do not specify here, although it is most
likely due to the antiferromagnetic fluctuations with a peak
around the momentum transfer $\vec Q=(\pi,\pi)$
\cite{Mazin08,Christianson08,Bang08,comment1}.

The simple result of $|\Delta_1| =|\Delta_2|$ is an accidental
consequence of the simple parameterization of
$\lambda_{11}=\lambda_{22}$ and $\lambda_{12}=\lambda_{21}$. For
more realistic parameterizations of the pairing interaction, the
magnitudes of the two gaps are different and additional peaks
show up in LDOS as shown in Fig.\ 2 below. One of many interesting
consequences of the $s\pi$ state is that the magnitude of one gap
in general is larger than the BCS value while the other gap is
smaller, that is, $2|\Delta_1|/T_c
> 3.52$ and $2|\Delta_2|/T_c < 3.52$. Another, perhaps more
interesting feature of the $s\pi$ pairing state is the appearance
of the zero bias peak which may be probed experimentally by the
tunneling spectroscopy. Szabo $et~al.$ recently performed
directional point contact Andreev reflection spectroscopy on
(Ba$_{0.55}$K$_{0.45}$)Fe$_2$As$_2$ and found that some of the
$ab$ plane spectra reveal the zero-bias conductance peak
consistent with the present work \cite{Szabo08}.

$LDOS~of ~S/N~ bilayers$ -- Now, consider the S/N bilayers of the
$s\pi$ pairing state described by Eq.\ (\ref{bdg2}) of the
thickness $y_t=y_S+y_N$. The Hamiltonian is written as an
$M\times M$ matrix, where $M=4y_S +2y_N$, on the basis of $\Psi$,
which is taken as
 \ba
\Psi_k^\dag = \left(
c_{1k\ua}^\dag(1),c_{1,-k\da}(1),c_{2k\ua}^\dag(1),c_{2,-k\da}(1),~~~~~
\right. \\ \nonumber
 \left. c_{1k\ua}^\dag(2),c_{1,-k\da}(2), \cdots, a_{k\ua}^\dag(y_t),a_{k\da}(y_t) \right).
 \ea

We first took the simple parameterization of $\lambda=0$,
$\lambda'=-0.8$, $t_0=t_1=t_2=0.25$ in the unit of the Fermi
energy $E_F$. With the parameterization, we diagonalized the
$M\times M$ matrix and calculated the gap function using Eq.\
(\ref{delta}). This procedure was repeated until the
self-consistency was reached. In Fig.\ 1(a) we show the zero
temperature 3 dimensional perspective plot of the LDOS of an S/N
bilayer of 20 S layers and 20 N layers as a function of energy in
the unit of the bulk pairing amplitude, $V/\Delta_0$. The ratio
of the gap to the Fermi energy in bulk, $\Delta_0/E_F$, is 0.053.
The coherence length in the unit of inter-layer distance is then
$\xi_S \approx 3-4 $ and the thickness of S layer is large enough
to monitor the evolution of the proximity effects as one moves
away from the S/N interface. Notice the LDOS enhancement in the
subgap energy region near the interface. The origin of this zero
bias enhancement is the sign change of the order parameter. This
point becomes clearer when we compare the present results with a
two gap superconductor of the same sign.

\begin{figure}[hbt]
 \epsfxsize=7cm \epsffile{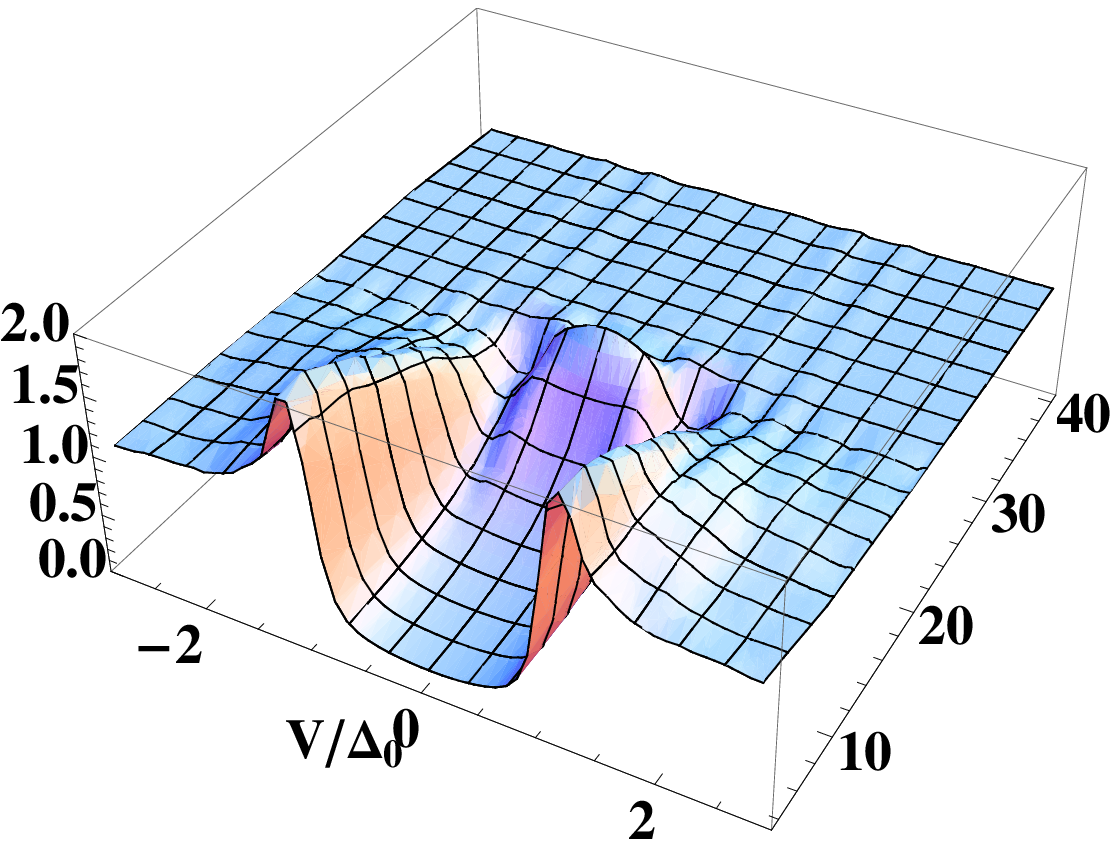} \epsfxsize=7cm \epsffile{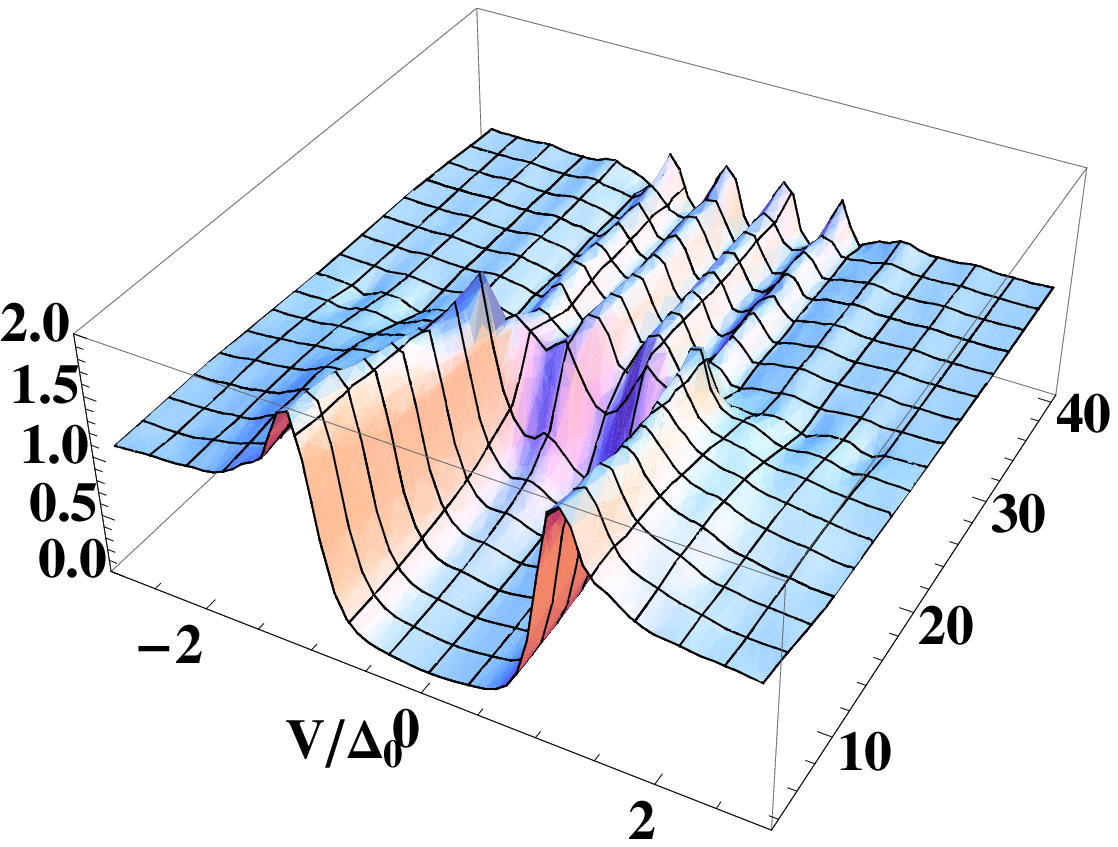}
\caption{(a) A 3D plot of LDOS of an $s\pi$ state S/N bilayer with
$\Delta_1 = -\Delta_2$ as a function of $V/\Delta_0$ and the
layer index. Notice the enhancement of the DOS in the subgap
region near the interface due to the sign reversal of the order
parameter. (b) The same as (a) but with $\Delta_1 = \Delta_2$. The
LDOS behaves exactly as the well established conventional S/N
bilayers.}\label{fig:SNm}
\end{figure}

For comparison we repeated the same calculations as (a) with all
the parameters remain unchanged except that the interband pairing
interaction is turned to a positive $\lambda'=0.8$, corresponding
to the case of Eq.\ (\ref{plus}). In this case,
$\Delta_1=\Delta_2$, and the LDOS should show the usual S/N
behavior as shown in Fig.\ 1(b). The expected proximity effects
are seen in the S and N regions. Compare this with the much short
ranged proximity effects in the N layers of Fig.\ (a). The
proximity effects of $\Delta_1$ and $\Delta_2$ for the $s\pi$
pairing state cancel each other almost exactly other than the
zero bias enhancement around the interface of S/N bilayers
because they have the opposite sign and equal amplitude.

The subgap enhancement is a robust feature of an $s\pi$ state. It
is a manifestation of the phase shift of $\pi$ between the two
condensates which is insensitive to parameters. To demonstrate
this we show in Fig.\ 2 the results of more realistic
parameterization. We took $\lambda_{11}=0.1,~ \lambda_{22}=0.2$,
$\lambda_{12}=-1.0, ~\lambda_{21}=-0.5$, $t_0=0.25$, and
$t_1=t_2=0.2$. As discussed above, other than another gap feature
shows up inside the larger gap, the zero bias enhancement can
clearly be seen. For the S/F $\pi$ state, the subgap enhancement
was observed for appropriate thickness of F where the order
parameter changes its sign, referred to also as DOS reversal
\cite{Kontos01}. For the $d$-wave pairing case, the sign change
of the order parameter occurs along the (110) surface, and the
zero bias conductance peak was also predicted and observed in the
cuprate superconductors \cite{Hu94,Wei98,Choi00}. For $s\pi$
state the peak appears unless one of $t_1$ and $t_2$ is negligible
such that both condensates are probed.

\begin{figure}[hbt]
 \epsfxsize=6cm \epsffile{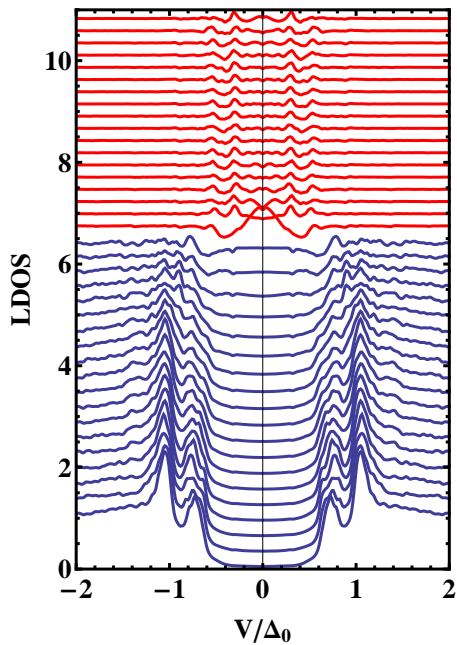}
\caption{ The LDOS of an $\pi$ state S/N bilayer with $|\Delta_1|
\ne |\Delta_2|$. The blue and red curves are LDOS in the S and N
layers, respectively. Each curve represents the LDOS of each
layer and is shifted upward for clarity. Notice the enhancement
of the DOS in the subgap region near the interface due to the
sign reversal of the order parameter. }\label{fig:SNp}
\end{figure}

$Summary ~and~ outlook$ -- We presented the local density of
states of an S/N bilayer based on a generic two band
superconductivity model with the internal $\pi$ phase. The sign
change of the pairing order parameter induced the LDOS enhancement
in the subgap region near the interface as was observed by
various tunneling spectroscopy on the FeAs superconductors. Some
of the experiments, however, exhibit much sharper conductance
peaks. This zero bias peak may be understood by more realistic
extensions of the present work. Dictated by the Hamiltonian of
Eq.\ (\ref{bdg2}), an electron of the intra-layer momentum $k$
tunneled from the N side may encounter the hole and electron band
pairing gaps with the amplitudes $t_1$ and $t_2$, respectively.
More realistically, however, the specularly reflected electron and
Andreev reflected hole may pick up the coherent phase difference
of $\pi$ if their intra-layer momenta fall on the electron and
hole Fermi surfaces, respectively. This process, which has been
included on an average manner in the present work, can produce a
much sharper conductance peak as observed in the pnictides and
along the (110) surface of the cuprates.

Other realistic considerations will be to allow the hopping
amplitudes of the electron and hole bands with the N layers, $t_1$
and $t_2$, to be different, or to consider more realistic pairing
interactions. When these extensions are included, the LDOS will
show more diverse and intriguing behavior. Also interesting will
be the spectroscopy of other kinds of $\pi$ state multilayers. For
instance, the S/F bilayers with the $s\pi$ state will exhibit
intriguing interplay between the internal and external $\pi$
phases. The works on these topics are in progress.

$Acknowledgement$ -- This work was supported by Korea Science \&
Engineering Foundation (KOSEF) through the Basic Research Program
Grant No.\ R01-2006-000-11248-0 (HYC) and by Korea Research
Foundation (KRF) through Grant No.\ KRF-2005-070-C00044 (HYC, YB)
and KRF-2007-521-C00081 (YB).


\begin{thebibliography}{99}
\bibitem{index} The zero bias peak due to the order parameter sign
change has a deep root founded on the Atiyah-Singer index theorem.
\bibitem{comment1} The sign reversal between the condensates means that the
interband pairing must be repulsive, that is,
$\lambda_{12},~\lambda_{21}<0$. This observation eliminates the
phonon and leaves the Coulomb repulsion of both charge and spin
channels for the pairing interaction. Together with this, the
observations that the two bands are separated by the $\vec
Q=(\pi,\pi)$ and the spin susceptibility has a peak also around
$\vec Q$ strongly hint that the antiferromagnetic fluctuations
act as the pairing glue.
\end{thebibliography}
\end{document}